\documentstyle[12pt]{article}
\input epsf
\begin{document}
\rightline{EFI-95-02}
\rightline{January 1995}
\rightline{hep-ph/9501291}
\vspace{0.7in}
\centerline{\bf CHARM AND BEAUTY IN PARTICLE PHYSICS\footnote{Presented at CERN
on September 28, 1994, at a symposium in honor of Andr\'e Martin's retirement.
This article is dedicated to the memory of M. A. Baqi B\'eg.}}
\vspace{0.5in}
\centerline{\it Jonathan L. Rosner}
\centerline{\it Enrico Fermi Institute and Department of Physics}
\centerline{\it University of Chicago, Chicago, IL 60637}
\vspace{0.5in}
\centerline{\bf ABSTRACT}
\medskip
\begin{quote}
The spectra of states containing charmed and beauty quarks, and their
regularities, are reviewed.
\end{quote}
\bigskip

\centerline{\bf I.  INTRODUCTION}
\bigskip

More than 20 years ago, two experimental groups announced the discovery of the
first in a series of charm-anticharm bound states \cite{Ting,psi}.  During the
first year in which the properties of these $c \bar c$, or {\em charmonium},
states were mapped out, this system began to display experimental possibilities
as rich as those in positronium.  However, an important difference from
positronium was predicted by theory and soon verified experimentally.  Whereas
the $2S$ and $1P$ positronium levels are nearly degenerate,\footnote{We label
levels by $n_r + 1$, where the radial quantum number $n_r$ is the number of
nodes of the radial wave function between 0 and $\infty$.} the $1P$ charmonium
level lies significantly below the $2S$ state.  What does this say about the
interquark force?  M. A. Baqi B\'eg asked this question of Andr\'e Martin
during Martin's visit to Rockefeller University in 1975. The result was the
first \cite{M77} in a series of lovely theorems about the order of energy
levels in nonrelativistic potentials \cite{Order,Erice,JS}, and a simple form
of power-law potential \cite{MPL} which has proved remarkably successful in
predicting the masses of new states containing not only charm and beauty, but
also strangeness.

My own involvement in similar questions began with the discovery of the upsilon
($b - \bar b$) levels \cite{ups}, for which the $2S - 1S$ spacing appeared
close to that in charmonium.  Chris Quigg and I asked what kind of potential
would give a level spacing independent of mass \cite{QR}.  The result, a
potential $V(r) \sim \ln r$ whose properties had been investigated even before
the discovery of the upsilons \cite{MT}, was surprisingly simple, and led us to
numerous related investigations of general properties of potential models
\cite{QRPR,Q79} and our own attempts at power-law fits \cite{GRR}.  It also
stimulated work in the inverse scattering problem \cite{TQR} as an outgrowth of
attempts to construct the interquark potential directly from data.

These parallel efforts have been marked by a good deal of correspondence
between the respective groups.  We have greatly enjoyed hearing about each
other's results.  It now appears that the first actual collaborative paper
involving both our groups \cite{PQN} will emerge as a result of this Symposium.
For this, and for the opportunity to honor Andr\'e, I am very grateful.

We begin in Section II by reviewing quarkonium spectra and their regularities.
We next discuss the predictions of power-law potentials for level spacings in
Sec.~III and for dipole matrix elements in Sec.~IV.  Some inverse scattering
results and the key role of information on the wave function at the origin are
mentioned in Sec.~V.  We discuss hadrons with one charmed quark in Sec.~VI, and
relate their properties to those of hadrons containing a single $b$ quark using
heavy quark symmetry in Sec.~VII.  An overview of the properties of hadrons
with beauty occupies Sec.~VIII.  These hadrons (particularly the mesons) are a
prime laboratory for the study of the Cabibbo-Kobayashi-Maskawa (CKM) matrix
(Sec.~IX) and of CP violation (Sec.~X). We note some issues for further study
and conclude in Sec.~XI.
\bigskip

\centerline{\bf II.  QUARKONIUM SPECTRA AND THEIR REGULARITIES}
\bigskip

Of all the known quarks, the charmed quark $c$ and the beauty quark $b$ offer
the best opportunity for the study of bound states and for insights into the
strong interactions using simple methods.  Since the scale at which the
interactions of quantum chromodynamics (QCD) become strong is several hundred
MeV, the masses of the $u$, $d$, and $s$ quarks are overwhelmed in bound states
by QCD effects.  The top quark is so heavy that it decays to $W + b$ before
forming bound states.  Leptons, of course, being colorless, do not participate
in this rich physics at all.  In this Section we give a brief overview of
levels containing only $c$ and $b$ quarks.
\bigskip

\leftline{\bf A.  Charmonium}
\bigskip

The charmonium spectrum is shown in Fig.~1.  Masses of observed levels are
based on the averages in Ref.~\cite{PDG}.  The prediction of the $\eta_c(2S)$
is based on Ref.~\cite{Msing}. Arrows are labeled by particles emitted in
transitions. States above the horizontal dashed line can decay to pairs of
charmed mesons ($D \bar D$) and are consequently broader than those below the
line, which decay both electromagnetically and with appreciable branching
ratios to non-charmed hadrons (not shown).

For many years, the major source of charmonium was the reaction $e^+ e^- \to
\gamma^* \to (c \bar c)$, which can produce only states with spin $J = 1$,
parity $P = -$, and charge-conjugation eigenvalue $C = -$, namely the $^3S_1$
and $^3D_1$ levels.  Other levels were reached by electric or magnetic dipole
transitions from the $J^{PC} = 1^{--}$ states, as indicated by the arrows
labeled by $\gamma$ in the figure.  More recently, starting with an experiment
in the CERN ISR \cite{ISR} and continuing with studies in the Fermilab
antiproton accumulator ring \cite{E760}, it has been possible to perform $\bar
p p$ collisions with carefully controlled energy, forming charmonium states in
the direct channel.  The observation of the $h_c(1P)$ level has been one
benefit of these studies, which are expected to continue.
\bigskip

\begin{figure}
\centerline{\epsfysize=4.6in \epsffile{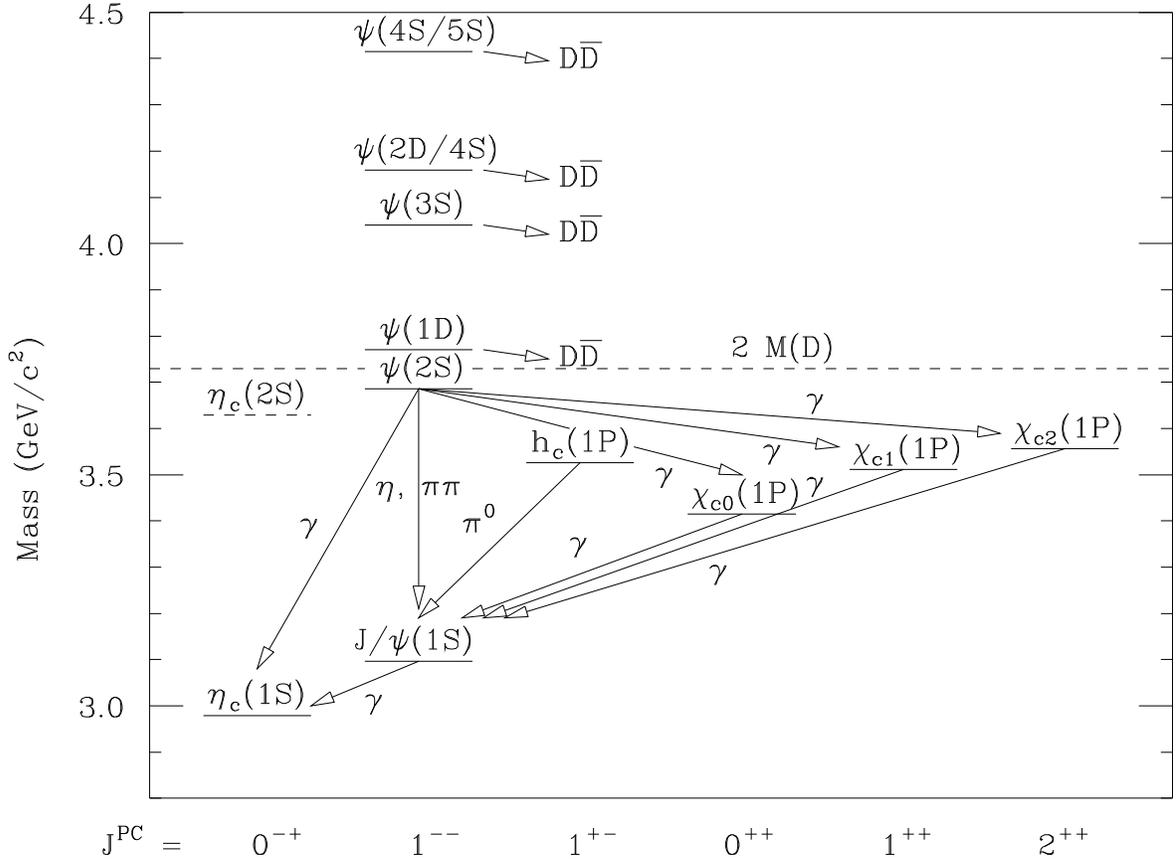}}
\caption{Charmonium ($c \bar c$) spectrum.  Observed and predicted levels are
denoted by solid and dashed horizontal lines, respectively.}
\end{figure}

\leftline{\bf B.  Upsilons}
\bigskip

We show $\Upsilon~(b \bar b)$ levels in Fig.~2.  The observed levels are as
quoted in Ref.~\cite{PDG}, while the $J^{PC} = 0^{-+}$ levels are shown with
masses predicted on the basis of Ref.~\cite{Msing}.  The $J^{PC} = 1^{+-}$
(``$h_b$'') levels are taken to have the spin-weighted average masses of the
corresponding $\chi_b$ levels. Since flavor threshold lies higher than for
charmonium, there are {\it two} sets of narrow P-wave levels, and consequently
a rich set of electric dipole transitions between the $\Upsilon$ and $\chi_b$
states, e.g., $3S \to 2P \to 2S \to 1P \to 1S$, $3S \to 1P$ (very weak), and
$2P \to 1S$. The systematics of these transitions has been a subject of recent
interest to Andr\'e, our colleagues, and me \cite{AMD,GRD}, which will be
described in Sec.~IV.
\bigskip

\begin{figure}
\centerline{\epsfysize=4.6in \epsffile{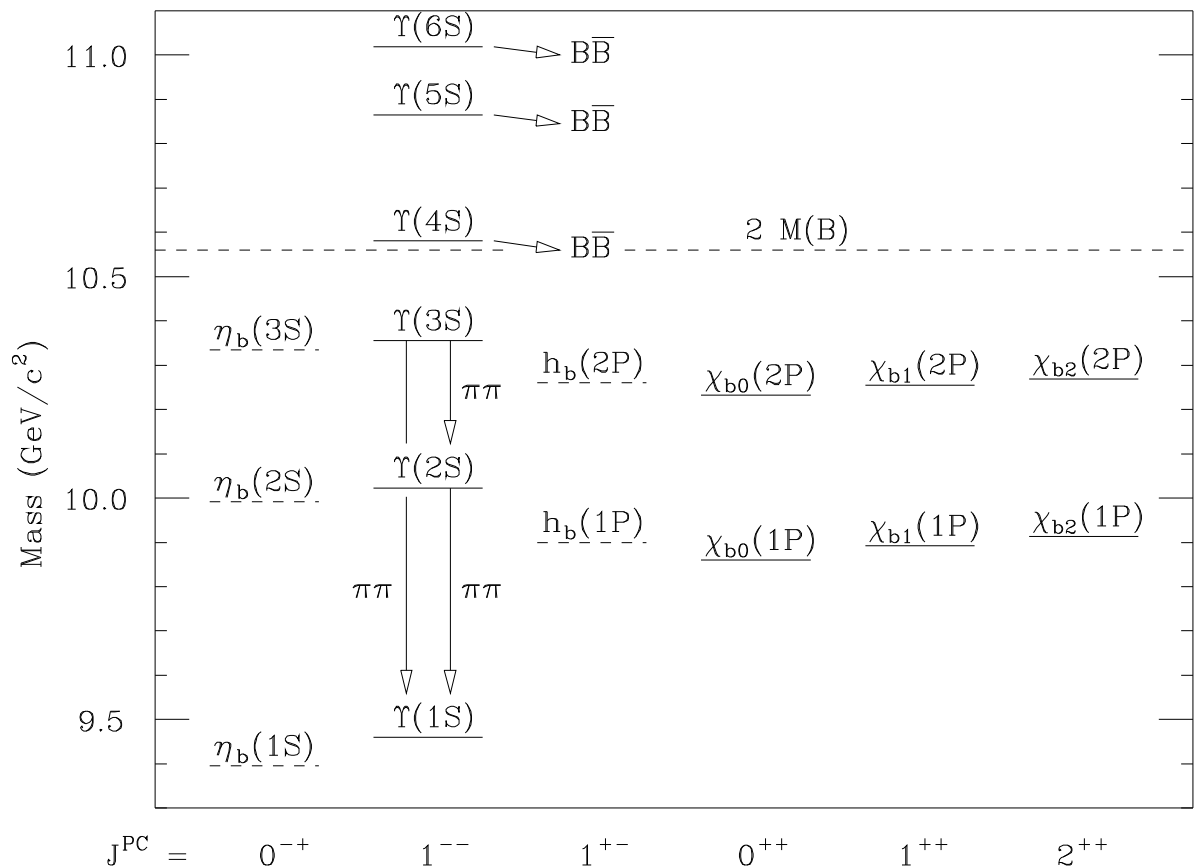}}
\caption{Spectrum of $b \bar b$ states.  Observed and predicted levels are
denoted by solid and dashed horizontal lines, respectively.  In addition to the
transitions labeled by arrows, numerous electric dipole transitions and decays
of states below $B \bar B$ threshold to hadrons containing light quarks have
been seen.}
\end{figure}

\leftline{\bf C.  Quarkonium and QCD}
\bigskip

As anticipated \cite{AP}, quarkonium has proved a remarkable laboratory
for the study of quantum chromodynamics.

{\it 1.  Forces} between a quark and an antiquark are best visualized with the
help of Gauss' Law.  At short distances, the interquark potential is described
by an effective potential $V(r) = - (4/3) \alpha_s(r)/r$, where the 4/3 is a
color factor and the strong fine structure constant $\alpha_s$ decreases as
$1/\ln r$ at short distances as a result of the asymptotic freedom of the
strong interactions \cite{AF}.  Lines of force behave approximately as they do
for a Coulomb potential.  They spread out in a typical dipole pattern; one
cannot tell the scale of the interaction by looking at them.  At long
distances, on the other hand, the chromoelectric lines of force bunch up into a
flux tube of approximately constant area, much as magnetic flux in a type-II
superconductor forms tubes.  The force between a quark and antiquark at long
distances is then independent of distance \cite{LD}, so the potential $V = kr$
rises linearly with distance.  Experimentally $k$ is about 0.18 GeV$^2$.

{\it 2.  Decays} of quarkonium states are a source of information about the
strength of the strong coupling constant.  For example, the ratio of the
three-gluon and $\mu^+ \mu^-$ decay rates of the $\Upsilon$ is proportional
to $\alpha_s^3/\alpha^2$, where $\alpha$ is the electromagnetic fine-structure
constant, and leads \cite{alphas} to a value of $\alpha_s(M_Z) = 0.108 \pm
0.010$ consistent with many other determinations.  (It has become conventional
to quote $\alpha_s$ at $M_Z$ even though the decay of the $\Upsilon$ probes
$\alpha_s$ at $m_b \simeq 5$ GeV.)

{\it 3.  Lattice QCD} calculations \cite{latal} deduce the value of $\alpha_s$
from the observed $1P - 1S$ level spacing in the $\Upsilon$ system (Fig.~2),
leading to $\alpha_s(M_Z) = 0.110 \pm 0.006$.  Both this value and that
determined from $\Upsilon$ decays are consistent with the world average
\cite{alphav} $\alpha_s(M_Z) = 0.117 \pm 0.005$.
\bigskip

\centerline{\bf III.  LEVEL SPACINGS IN POWER-LAW POTENTIALS}
\bigskip

The spectra of the Coulomb ($V \sim - r^{-1}$) and three-dimensional oscillator
($V \sim r^2$) potentials are familiar to students of quantum mechanics, some
of whom even are aware (as was Newton \cite{Chandra,clasorb}) that
the two problems are related to one another.  These spectra are illustrated
in Figs.~3(a) and 3(b).  In the Coulomb potential, the energy levels are
proportional to $-(n_r + L + 1)^{-2} = - n^{-2}$, where $n$ is the principal
quantum number, and thus are highly degenerate.  A different type of degeneracy
is present in the harmonic oscillator, for which the energies are proportional
to $2n_r + L + 3$.  An intermediate case, $V \sim \ln r$ (equivalent to the
limit of $V = (r^\nu - 1)/\nu$ as $\nu \to 0$) is shown in Fig.~3(c).  (Further
examples may be found in Ref.~\cite{PQN}).  The quarkonium spectrum is rather
similar to this.  Indeed, a potential $V(r) = - (4/3) \alpha_s(r)/r + kr$ can
be approximated by some power intermediate between $-1$ and 1 for a limited
range of distance \cite{HJL}.  It so happens that for $c \bar c$ and $b \bar b$
states, which are sensitive to the range between 0.1 and 1 fm \cite{Q79}, this
power turns out to be close to zero.
\bigskip

\begin{figure}
\centerline{\epsfysize=6in \epsffile{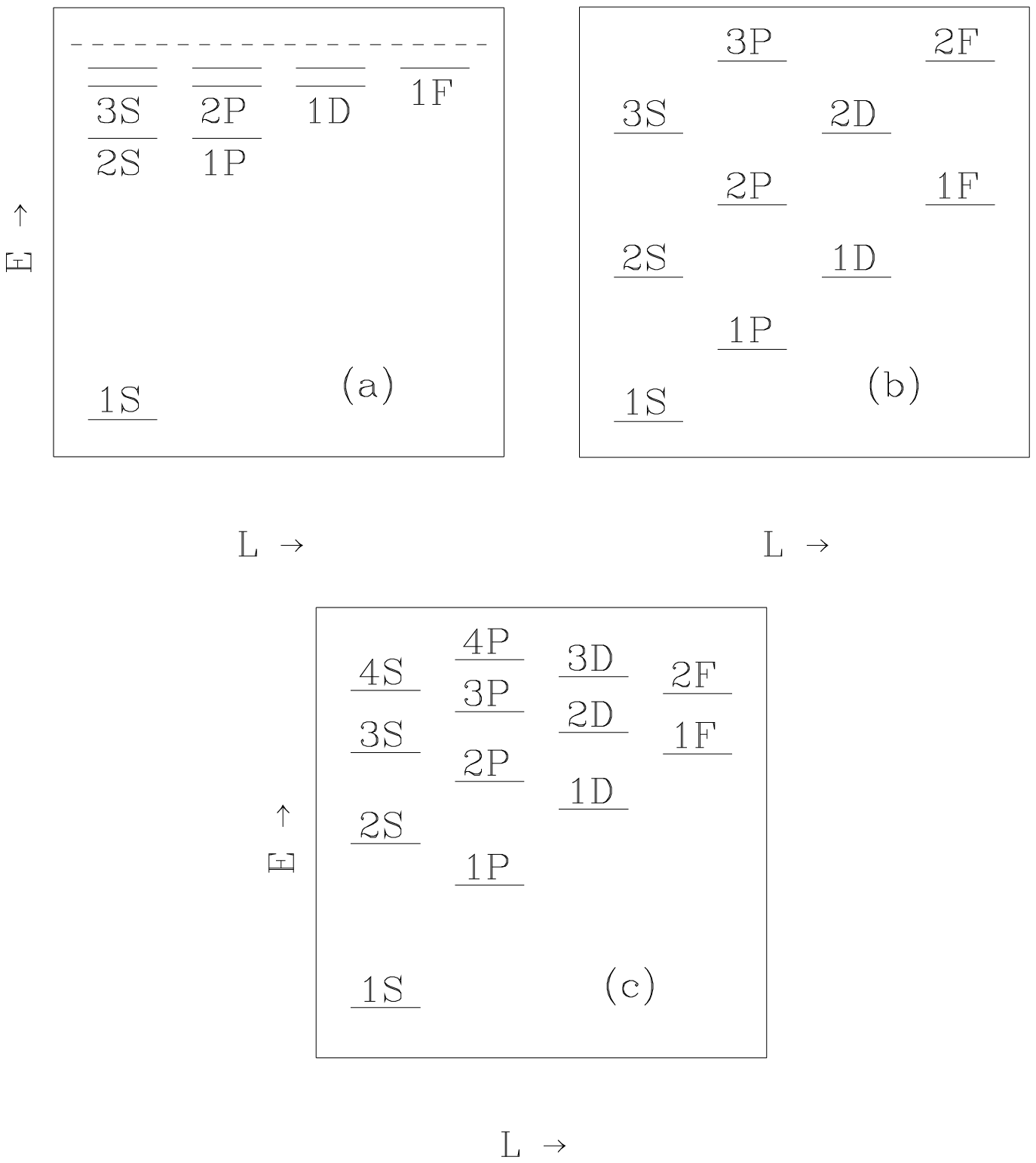}}
\caption{Patterns of lowest-lying energy levels in various potentials $V(r) =
(r^\nu - 1)/\nu$. (a)  Coulomb potential ($\nu = -1$) (the dashed line
indicates the onset of continuum levels); (b) three-dimensional oscillator
($\nu = 2$); (c) $V(r) \sim \ln r$, corresponding to the limit $\nu \to 0$.}
\end{figure}

\leftline{\bf A.  Predictions of the Martin potential}
\bigskip

The $2S - 1S$ level spacing in the $\Upsilon$ family is slightly smaller than
that in charmonium.  Since level spacings in a potential $V \sim r^\nu$ behave
with reduced mass $\mu$ as $\Delta E \sim \mu^{-\nu/(\nu+2)}$ \cite{QR,QRPR}, a
small positive power will be able to reproduce this feature.  What is
remarkable is how much else is fit by such a simple ansatz. A potential $V(r)
\sim r^{0.1}$ \cite{MPL} (we refer the reader to the original articles for
precise constants and quark masses) not only fits charmonium and upsilon
spectra remarkably well, as shown in Fig.~4, but also has been successful in
fitting and anticipating masses of states containing strange quarks, using the
mass of the $\phi(1020)$ as the input for the $1^3S_1~ s \bar s$ level.  We
compare these predictions with observations \cite{PDG,CLEODss,Sharma,CUSBBs}
in Table 1.  Standard assumptions
regarding spin-spin interactions have been made in order to estimate hyperfine
splittings between $^1S$ and $^3S$ levels.
\bigskip

\begin{figure}
\centerline{\epsfysize=5in \epsffile{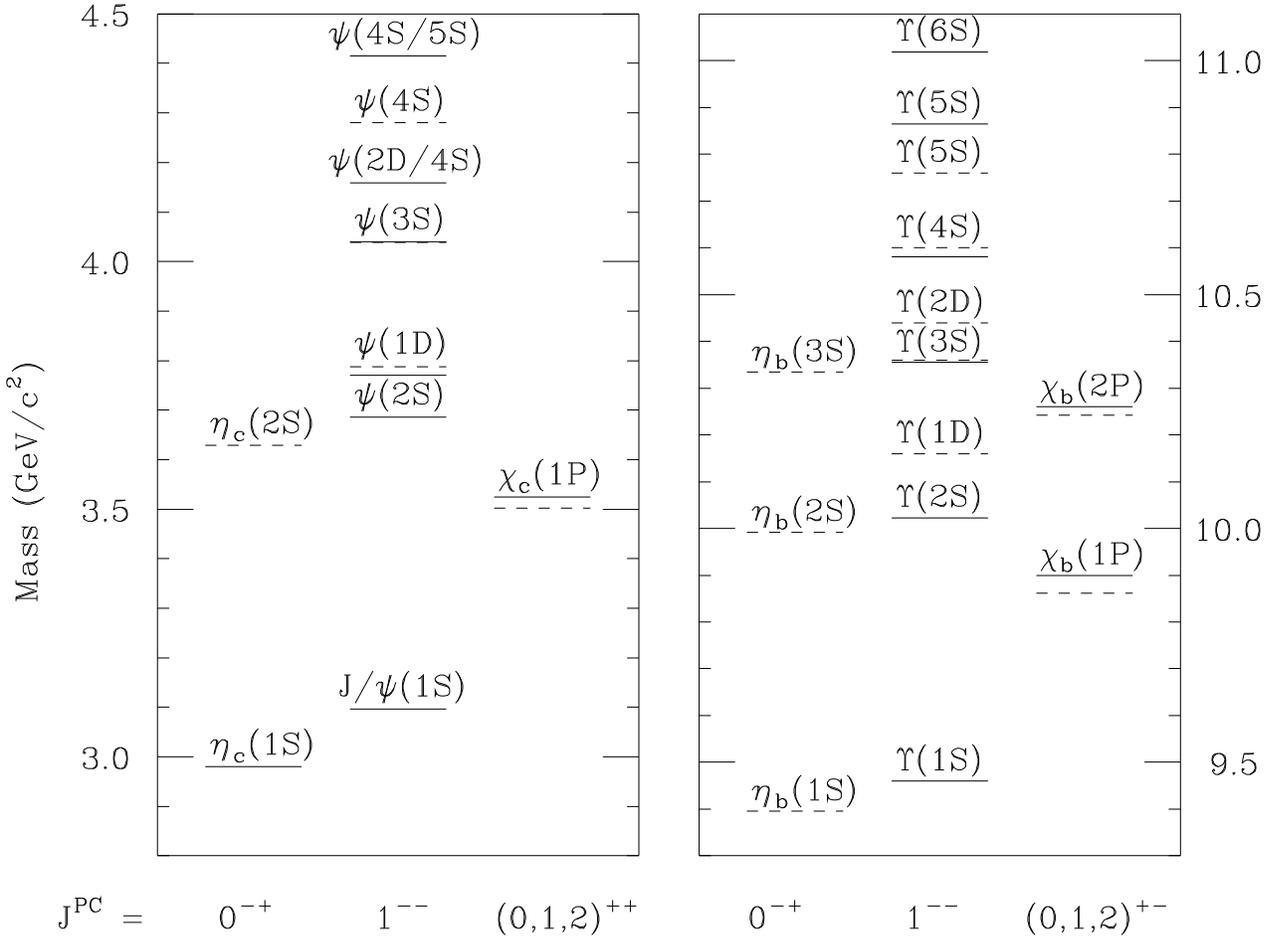}}
\caption{Comparisons of prediction of Martin's potential with experiment for
(a) charmonium and (b) upsilon levels.  Solid lines denote experimental values;
dashed lines denote predictions (where noticeably different from
observations).}
\end{figure}

\begin{table}
\caption{Masses of states containing strange quarks predicted in a potential
$V \sim r^\nu$ and observed experimentally.}
\begin{center}
\begin{tabular}{|c c c|c c c|} \hline
Level & Predicted   & Observed   & Level & Predicted  & Observed   \\
      & Mass (GeV)  & Mass (GeV) &       & Mass (GeV) & Mass (GeV) \\ \hline
$(s \bar s)_{2S}$ & 1.634 & 1.650$^{a)}$ & $b \bar s$ & 5.364 & 5.368$^{d)}$ \\
                  &       &            &       & $\pm 0.010$ & $\pm 0.004$ \\
$c \bar s$ & 1.99 & 1.97  & $(b \bar s)^*$ & 5.409 & $5.422^{e)}$ \\
                  &       &            &       &             & $\pm 0.006$ \\
$(c \bar s)^*$ & 2.11 & 2.11  & $b \bar c$ & 6.25  &    \\
$(c \bar s)^{~3}P$ & 2.54 & 2.54$^{b)}$ & $(b \bar c)^*$ & 6.32 & \\
                &      & 2.57$^{c)}$ &               &      & \\ \hline
\end{tabular}
\end{center}
\leftline{$^{a)}$ Ref.~\cite{PDG}; $^{b)~3}P_1$ level \cite{PDG};
$^{c)~3}P_2$ level \cite{CLEODss};}
\leftline{$^{d)}$ Ref.~\cite{Sharma}; $^{e)}$ Ref.~\cite{CUSBBs}; see
discussion in text.}
\end{table}

\leftline{\bf B.  Remarks on levels}
\bigskip

The agreement between predictions and experiment in Fig.~4 and Table 1 is so
good that many predictions are hard to distinguish from the observations.  Even
the discrepancies are interesting.

{\it 1.  The $\eta'_c$}, when predicted, disagreed with a claimed state
\cite{oldetac} which has not been confirmed in a new proton-antiproton
experiment \cite{E760}.

{\it 2.  The observed $\psi(1D)$ level}, the $\psi(3770)$, is a $1^3D_1$ state,
whereas the prediction has been shown for the spin-averaged $1D$ mass.  The
other $1D$ levels (the $^{1,3}D_2$ and $^3D_3$) probably lie higher, and are
accessible in $\bar p p$ interactions.  The $\psi(3770)$ is a good source of
$D \bar D$ pairs, soon to be exploited by the Beijing electron-positron
collider.  The $^{1,3}D_2$ levels cannot decay to $D \bar D$ and probably
lie below $D \bar D^*$ threshold, so they are expected to be narrow.

{\it 3.  The observed $\psi(4160)$ level} is not really understood on the basis
of {\em any} simple potential models, Martin's or otherwise. Is it the $2^3D_1$
level, mixed with S-waves so as to have an appreciable coupling to $e^+ e^-$?
Its mass and couplings are undoubtedly strongly affected by coupled channels.
A similar distortion is visible near $B \bar B$ threshold in the $\Upsilon$
family \cite{BE}.

{\it 4.  The $\chi_b(9900)$ levels} lie higher than Martin's prediction,
exposing the limitations of a universal power-law potential.  Their position
relative to the $1S$ and $2S$ levels, when compared to that of the $\chi_c$
levels in charmonium, is weak evidence that the interquark potential is
becoming more singular at short distances, as predicted by QCD \cite{KQR}.

{\it 5.  The $1D$ and $2D$ $b \bar b$ levels} can be searched for in the
direct $e^+ e^- \to \gamma^* \to ~^3D_1$ reaction, in cascade reactions
involving electric dipole transitions to and from P-wave levels, and
possibly in transitions to $\Upsilon(1S) \pi \pi$ \cite{Ko}.

{\it 6.  The $D_s^* - D_s$ splitting} is about the same as the $D^* - D$
splitting.  Since the hyperfine splitting is proportional to $|\Psi(0)|^2/m_c
m_q$, where $\Psi(0)$ is the nonrelativistic wave function of the charmed quark
and the light quark $q = d,~s$ at zero separation, one expects $|\Psi(0)|^2_{c
\bar s} \approx |\Psi(0)|^2_{c \bar d} (m_s/m_d)$, a relation useful in
determining the ratio of the $D_s^+$ and $D^+$ decay constants \cite{JRFM}.

{\it 7.  The $B_s^* - B_s$ splitting} in Martin's approach, as well as in an
expansion in inverse powers of heavy quark masses performed much later
\cite{RW}, is predicted to be the same as the $B^{*0} - B^0$ splitting. A
tentative observation by the CUSB group \cite{CUSBBs} is consistent with this
expectation.

{\it 8.  The ratio of level spacings $(3S - 2P)/(2P - 1D)$} is an interesting
quantity.  In a power-law potential $V \sim r^\nu$, for a wide range of values
of $\nu$, this quantity is expected to be very close to unity \cite{PQN}. This
circumstance can be useful to anticipate the position of the $b \bar b$ $1D$
levels, discussed above, and the $c \bar c$ $2P$ levels, which may play a role
\cite{WiseClose} in the hadronic production of the $\psi(2S)$ \cite{CDFpsi}.
This ratio for $b \bar b$ states is very far from unity in Ref.~\cite{Erice},
where Martin quoted a prediction for the $1D$ levels from another source
\cite{KR}.

{\it 9.  The $^1P_1$ levels of quarkonium} were predicted by Stubbe and Martin
\cite{SM} to lie no lower than the spin-weighted average of the corresponding
$\chi(^3P_{0,1,2})$ levels.  A candidate for the $^1P_1$ $b \bar b$ level
proposed by the CLEO Collaboration several years ago \cite{CLEOh} violated this
bound; it was subsequently not confirmed.  The corresponding $c \bar c$ level
has been discovered just at the lower limit of the Stubbe-Martin bound
\cite{hc}; its mass is $3526.14 \pm 0.24$ MeV, close to the spin-weighted
average of the $\chi_c$ levels, $3525.3 \pm 0.1$ MeV.
\bigskip

\leftline{\bf C.  Mesons with charm and beauty}
\bigskip

An interesting system in which the quarks are heavy but unequal in mass is the
set of $b \bar c$ levels, recently discussed in detail by Eichten and Quigg
\cite{EQ}. The positions of their predicted $1S$ levels are very close to those
anticipated by Martin (see Table 1).  If the fine structure of the $1P$ levels
(predicted to lie around 6.7 GeV) can be observed, it may provide new
information about spin-dependent forces not accessible in equal-mass systems.
The $2S - 1S$ spacing is predicted to be somewhat below 0.6 GeV. A narrow set
of $1D$ levels is predicted at 7.0 GeV. The $2P$ levels are expected to lie
very near the $\bar B D$ threshold at 7.14 GeV.
\bigskip

\centerline{\bf IV.  DIPOLE TRANSITIONS IN POWER-LAW POTENTIALS}
\bigskip

The pattern of electric dipole matrix elements in atomic transitions can
be understood on very intuitive grounds, in terms of overlaps of wave
functions and semiclassical arguments \cite{olddip}.  The $\Upsilon$
system is rich enough to display some aspects of this pattern, as shown in
Fig.~5 \cite{GRD}.

\begin{figure}
\centerline{\epsfysize=3.5in \epsffile{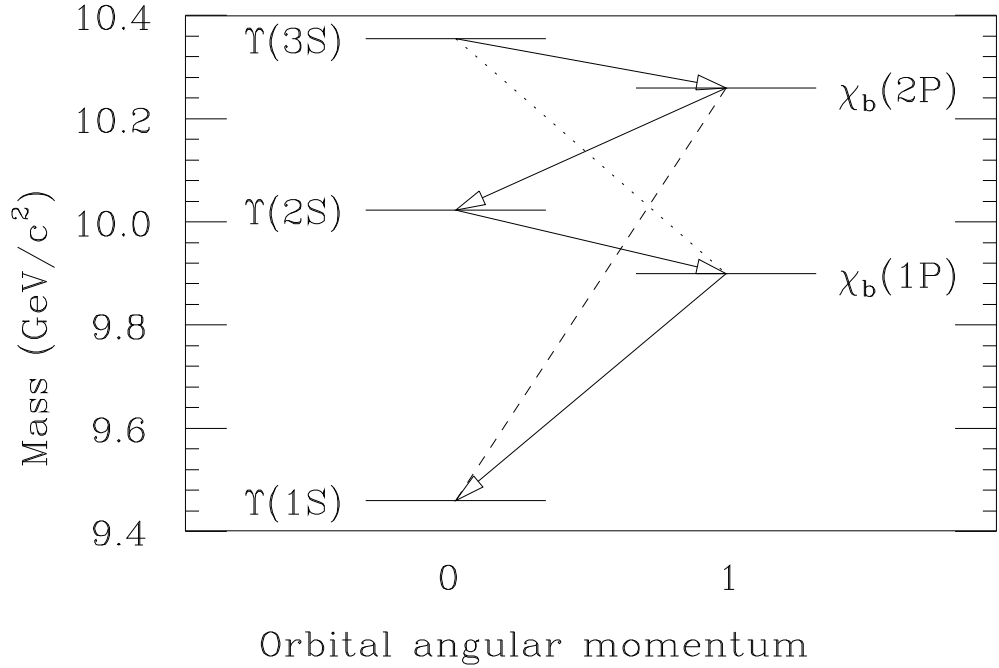}}
\caption{Observed electric dipole transitions in the $\Upsilon$ system.
Arrows denote favored transitions.  The very weak $3S \to 1P$ transition is
denoted by a dotted line.  The $2P \to 1S$ transition, denoted by a dashed
line, is also somewhat suppressed.}
\end{figure}

\begin{figure}
\centerline{\epsfysize=3.5in \epsffile{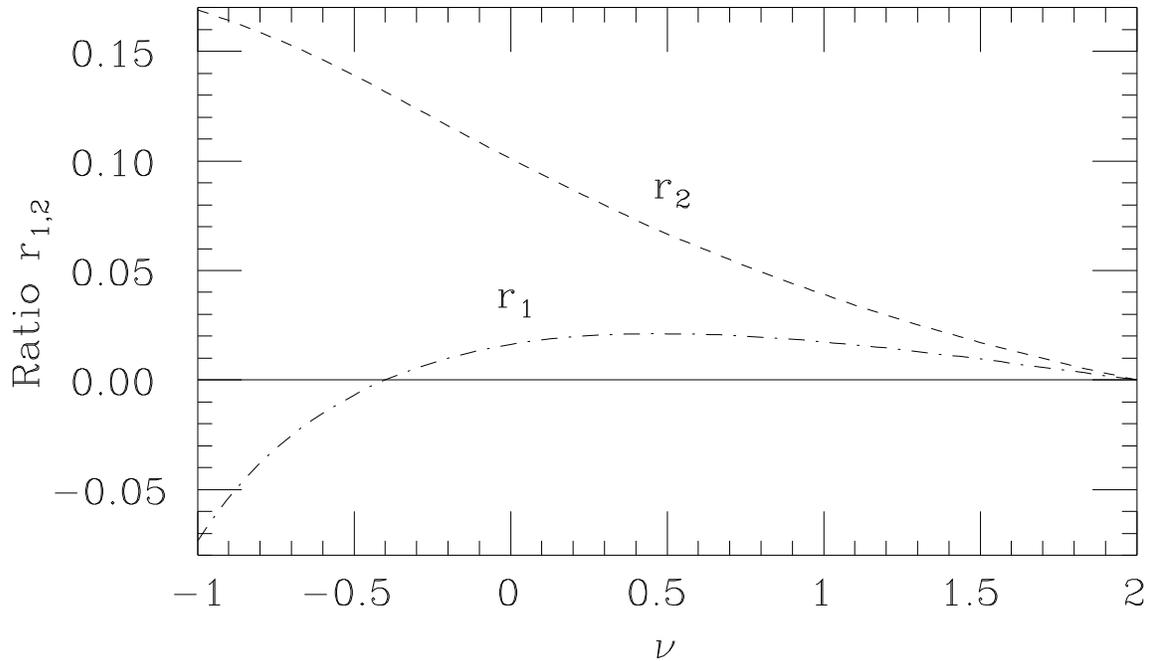}}
\caption{Ratios $r_1 \equiv \langle 1P | r | 3S \rangle / \langle 2P | r | 3S
\rangle$ (dot-dashed) and $r_2 \equiv \langle 1S | r | 2P \rangle / \langle 2S
| r | 2P \rangle$ (dashed) as a function of $\nu$ in power-law potentials $V(r)
\sim r^\nu$.}
\end{figure}

Let us denote the orbital angular momentum by $L$, the radial quantum
number by $n_r$, and the principal quantum number by $n = n_r + L + 1$.  (We
have been labeling our levels by $n_r + 1$.) As in atoms, transitions in which
$n$ and $L$ change in opposite directions are highly disfavored. For
example, in the transition $3S \to 1P$, $n$ decreases from 3 to 2 while $L$
increases from 0 to 1.  Such transitions are just barely visible in the
$\Upsilon$ system \cite{CUSBdip}.  The ratio $ r_1 \equiv \langle 1P | r | 3S
\rangle / \langle 2P | r | 3S \rangle$ is highly suppressed in power-law
potentials for a large range of interesting powers, as seen in Fig.~6.

There is also a tendency for transitions to favor levels whose wave functions
are as similar to one another as possible.  Thus, the transition $2P \to 1S$
(involving a change of two units of $n$) is suppressed in comparison with $2P
\to 2S$, where $n$ changes by only one unit.  Fig.~6 shows that the ratio $ r_2
\equiv \langle 1S | r | 2P \rangle / \langle 2S | r | 2P \rangle$ is moderately
suppressed in power-law potentials. Both $r_1$ and $r_2$ would vanish in a
harmonic oscillator potential, as can be seen by expressing the dipole operator
as a sum of creation and annihilation operators.

While working on dipole transitions ~\cite{GRD}, we had enjoyable
correspondence with Andr\'e, who shared with us a number of interesting
rigorous results \cite{AMD} on the signs of dipole matrix elements in various
potentials.  A number of years ago, Andr\'e had already shown that the $2P \to
1S$ matrix element could not vanish and had the same sign as the product of the
two radial wave functions at infinity \cite{oldAMD}.
\bigskip

\centerline{\bf V.  INVERSE SCATTERING RESULTS}
\bigskip

One can construct an interquark potential directly from the masses and leptonic
widths of S-wave quarkonium levels \cite{TQR}.  A potential constructed from $b
\bar b$ levels agrees remarkably well with that constructed using charmonium
data, except at the shortest distances, where the heavier $b \bar b$ system
provides the more reliable information.  (We refer the reader to
Refs.~\cite{TQR} for illustrations.)  Consistency between the two constructions
leads to a rather tight constraint on the difference between charmed and $b$
quark masses, $m_b - m_c \simeq 3.4$ GeV.

Supersymmetric quantum mechanics \cite{Wit} has proved very helpful
in the construction of potentials \cite{SSQM}.  A Hamiltonian with a given
spectrum can be factorized into the product of two operators, $H_+ = A^{\dag}
A$.  A Hamiltonian $H_- = A A^{\dag}$ (related by supersymmetry to $H_+$) has
the same spectrum aside from any state $|0 \rangle$ annihilated by the operator
$A$, in which case $|0 \rangle$ is the (zero-energy) ground state of $H_+$, but
does not belong to the spectrum of $H_-$.  Starting from a potential $V_- =
\kappa^2$ in $H_-$ which has no bound states, we then find a potential $V_+ =
\kappa^2[1 - 2~{\rm sech}^2 \kappa(x - x_0)]$ in $H_+$ with a single
zero-energy bound state.  The integration constant $x_0$ may be chosen to give
a symmetric potential whose odd-parity levels are suitable S-wave wave
functions for the radial equation of a three-dimensional problem.  By
appropriate shifts of the energy after each supersymmetry transformation, one
can construct potentials with an arbitrary spectrum.  This construction bears
an interesting relation to the vertex operator in string theory \cite{GRSSQM}.

The key role of leptonic widths in solving the inverse scattering problem
arises from the information they provide on the squares of wave functions
at zero interquark separation.  These quantities obey beautiful regularities
and inequalities in power-law potentials \cite{QRPR,MLW}.
\newpage

\centerline{\bf VI.  CHARMED HADRONS}
\bigskip

\begin{figure}
\centerline{\epsfysize=4.7in \epsffile{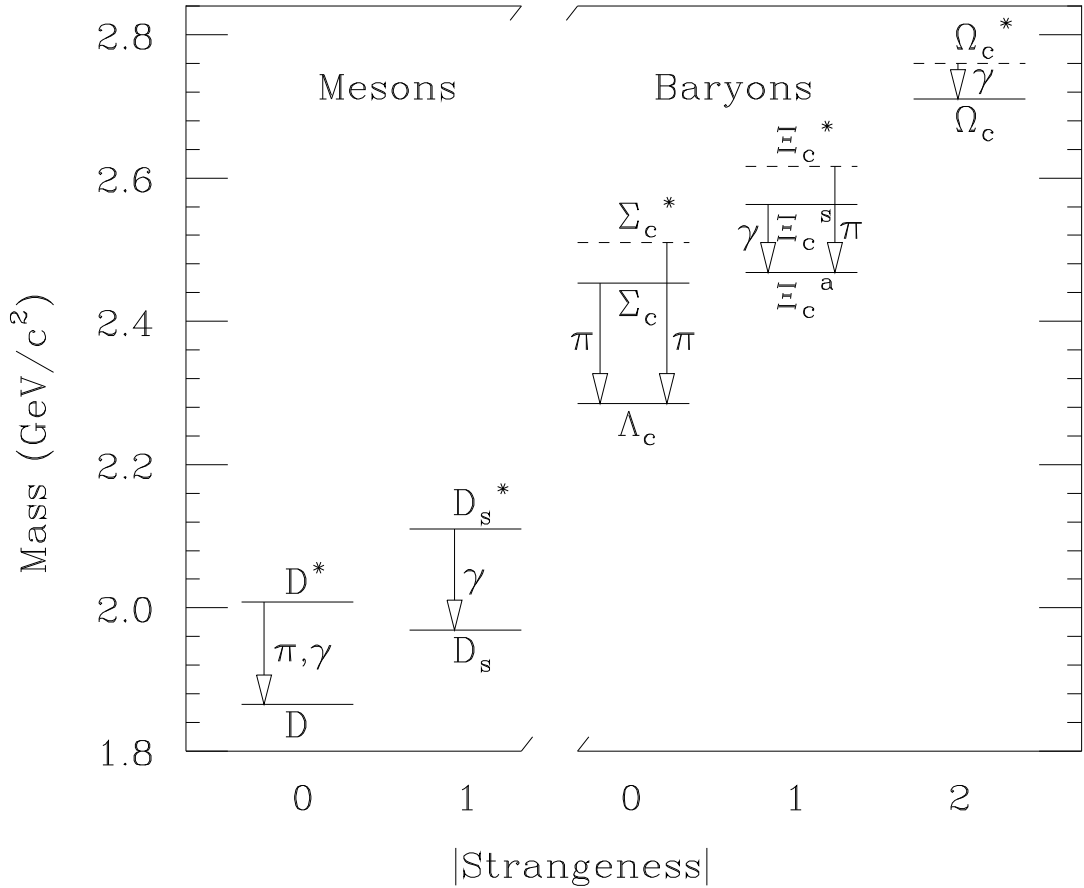}}
\caption{Lowest-lying S-wave levels of hadrons containing a single charmed
quark.  The lowest level in each group decays weakly.  Dashed lines indicate
levels not yet observed.}
\end{figure}

The ground states of levels containing a single charmed quark are shown in
Fig.~7, adapted from Ref.~\cite{KQR} using data quoted in Ref.~\cite{PDG}.  All
the levels except baryons with spin 3/2 (dashed lines) have been seen,
including a recently reported excited state of the $\Xi_c$ found in a CERN
experiment \cite{Xics}. What follows is a small sample of some interesting
questions in charmed-hadron physics.
\bigskip

\leftline{\bf A.  $D$ meson semileptonic decays}
\bigskip

A free-quark model of $D$ meson semileptonic decays in which the charmed quark
undergoes the transition $c \to s \ell^+ \nu_\ell$ would predict, in the limit
of zero recoil of the strange quark, the ratio of 1:3:0 for $\bar K:\bar
K^*:\bar K^{**}$, where $\bar K^{**}$ stands for any excited state of the
strange quark and nonstrange spectator antiquark.  The observed ratio is more
like 7:4:(0 to 4) \cite{PDG,Bean,AR};
it is still not certain how much of the $D$ semileptonic
branching ratio is associated with states other than $\bar K$ and $\bar K^*$.
($B$ meson semileptonic decays lead to final states other than $D$ and $D^*$
\cite{StSt}, so one should expect similar behavior for lighter-quark
systems.)

Jim Amundson and I have looked at this process \cite{AR} from the standpoint of
heavy quark effective theory, treating the strange quark as heavy in a manner
reminiscent of Andr\'e's bold assumption for quarkonium spectra, mentioned in
Sec.~III.  We can identify several sources of the discrepancy with the
heavy-quark limit, including an overall QCD suppression of $\bar K$ and
$\bar K^*$ production, a phase-space suppression of $\bar K^*$ relative to
$\bar K$, and a spin-dependent (hyperfine) interaction between the strange
quark and the spectator antiquark which increases the $\bar K$ rate and
decreases the $\bar K^*$ rate.
\bigskip

\leftline{\bf B.  Strange $D$ meson decay constants}
\bigskip

Recent observations of the decay $D_s \to \mu \nu$ \cite {Dmunu} have led to a
measurement of the quantity $f_{D_s} \simeq 300$ MeV (in units where the pion
decay constant is 132 MeV).  This value agrees with one obtained earlier
\cite{JRFM,ArgDs} from the decay $\bar B \to D_s^- D$ under the assumption that
the weak current in the decay of a $b$ quark to a charmed quark creates a
$D_s^-$ meson. Through the expression $f_{D_s}^2 = 12 |\Psi(0)|^2 / M_{D_s}$,
where $\Psi(0)$ is the wave function of the charmed quark and strange antiquark
at zero separation, and the use of heavy quark symmetry, one can extrapolate
this observation to predict other heavy meson decay constants, such as $f_D$,
$f_B$, and $f_{B_s}$.  A measurement of $f_D$ may be available in the near
future at the Beijing Electron-Positron Collider (see Sec.~III B 2). The last
two decay constants are of particular interest in the study of CP violation in
$B$ meson decays, as we shall see.
\bigskip

\leftline{\bf C.  Charmed baryons}
\bigskip

{\it 1.  Excited strange baryons} ought to be visible in semileptonic decays
of the $\Lambda_c$.  The nonstrange quarks in a $\Lambda_c$ are in a state of
spin and isospin zero.  In a spectator model, they should remain so.  If the
strange quark is given a sufficient ``kick,'' the nonstrange quarks should
be able to form not only a $\Lambda$, but also the lowest-lying excitations
in which the nonstrange quarks have zero spin and isospin, the states
$\Lambda(1405)$, with $J^P = 1/2^-$, and the $\Lambda(1520)$, with $J^P =
3/2^-$.  No such states have yet been seen \cite{Shipsey}; why not?

Many decays of $\Lambda(1405)$ and $\Lambda(1520)$ are elusive, consisting of
charged $\Sigma \pi$ modes, and $\bar K^0 n$ for the $\Lambda(1520)$.  However,
the decays $\Lambda(1405) \to \Sigma^0 \pi^0 \to \Lambda \gamma \pi^0$ and
$\Lambda(1520) \to K^- p$ are visible in CLEO. The importance of such final
states in semileptonic $\Lambda_c$ decays consists not only in the degree to
which semileptonic decays of heavy-quark hadrons populate excited states, but
in the normalization of numerous branching ratios of the $\Lambda_c$ \cite{DR}.

{\it 2.  Excited charmed baryons} have recently been identified \cite{EXLC},
consisting of states lying 308 and 342 MeV above the $\Lambda_c$. Since the
light-quark system in a $\Lambda_c$ baryon consists of a $u$ and $d$ quark
bound to a state $[ud]$ of zero spin, zero isospin, and color antitriplet, the
$\Lambda_c$ is a simple object in heavy-quark symmetry, easily compared with
the $\Lambda_b = b[ud]$ and even with the $\Lambda = s[ud]$.

The $[ud]$ diquark in the $\Lambda$ can be orbitally excited with respect to
the strange quark.  The $L = 1$ excitations consist of a fine-structure
doublet, the $\Lambda(1405)$ with spin-parity $J^P = 1/2^-$ and the
$\Lambda(1520)$ with $J^P = 3/2^-$ mentioned above.  The spin-weighted average
of this doublet is 366 MeV above the $\Lambda$.  These states are illustrated
on the left-hand side of Fig.~8.

The candidates for the charmed counterparts of the $\Lambda(1405)$ and
$\Lambda(1520)$ are shown on the right-hand side of Fig.~8. The spin-weighted
average of the excited $\Lambda_c$ states is 331 MeV above the $\Lambda_c$, a
slightly smaller excitation energy than that in the $\Lambda$ system.  The
difference is easily understood in terms of reduced-mass effects.  The ${\bf L
\cdot S}$ splittings appear to scale with the inverse of the heavy quark ($s$
or $c$) mass. The corresponding excited $\Lambda_b$ states probably lie 300 to
330 MeV above the $\Lambda_b(5630)$, with an ${\bf L \cdot S}$ splitting of
about 10 MeV.
\bigskip

\begin{figure}
\centerline{\epsfysize=3.25in \epsffile{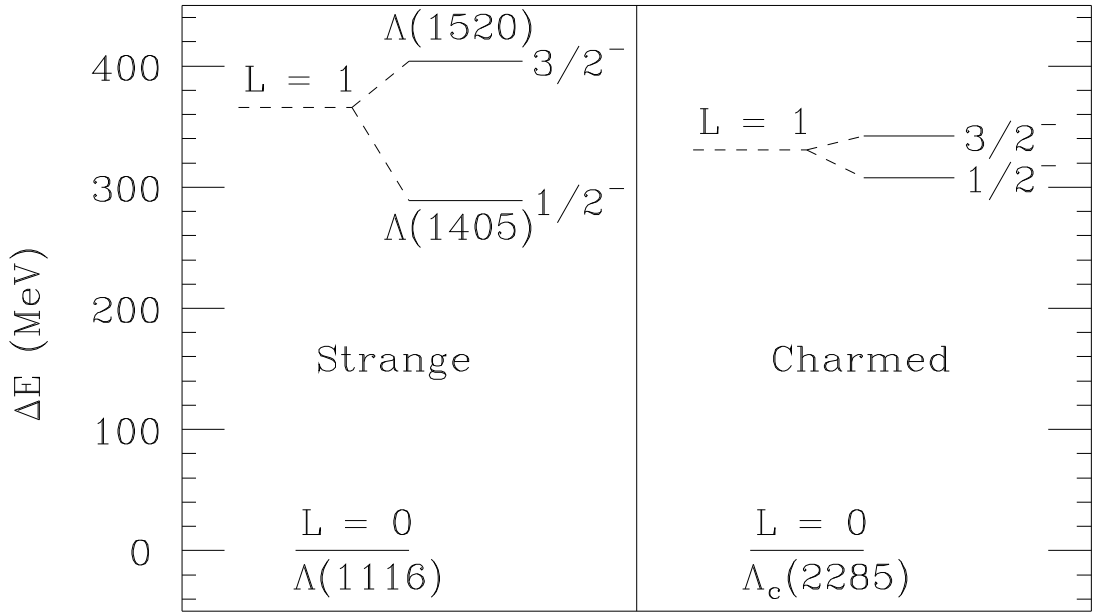}}
\caption{Ground states and first orbital excitations of $\Lambda$ and
$\Lambda_c$ levels.}
\end{figure}

\leftline{\bf D. Excited charmed mesons}
\bigskip

A good deal of progress has been made recently in the study of the P-wave
resonances of a $c$ quark and a $\bar u$ or $\bar d$, generically known as
$D^{**}$ states.  Present data \cite{PDG,CLEODss} and predictions \cite{EHQ}
are summarized in Fig.~9.

\begin{figure}
\centerline{\epsfysize=3in \epsffile{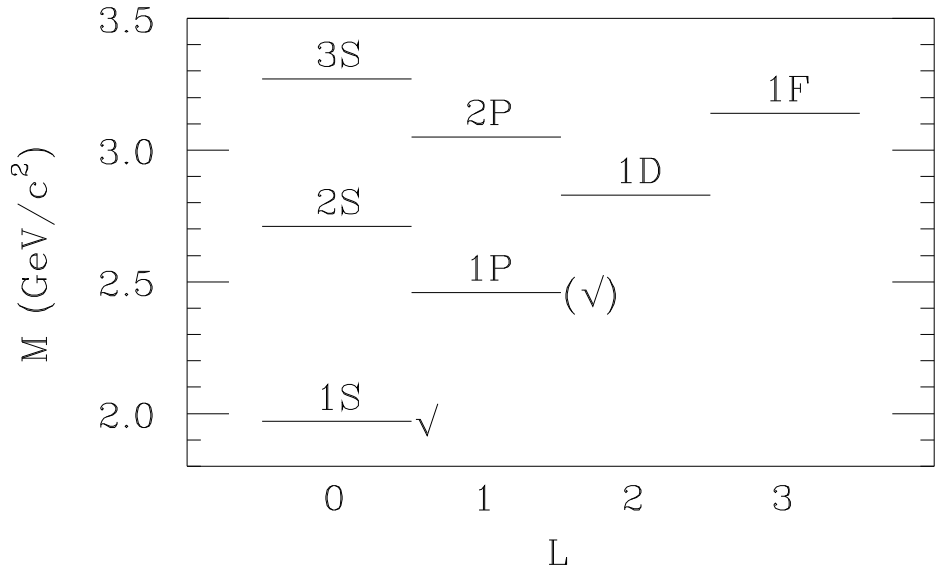}}
\caption{Low-lying nonstrange resonances of a $c$ quark and a light ($\bar u$
or $\bar d$) antiquark.  Check marks with or without parentheses denote
observation of some or all predicted states.}
\end{figure}

The observed states consist of the $1S$ (singlet and triplet) charmed mesons
and all six (nonstrange and strange) $1P$ states in which the light quarks'
spins combine with the orbital angular momentum to form a total light-quark
angular momentum $j = 3/2$. These states have $J = 1$ and $J = 2$.  They are
expected to be narrow in the limit of heavy quark symmetry.  The strange $1P$
states are about 110 MeV heavier than the nonstrange ones. In addition, there
are expected to be much broader (and probably lower) $j = 1/2~D^{**}$
resonances with $J = 0$  and $J = 1$.

For the corresponding $B^{**}$ states, one should add about 3.32 GeV (the
difference between $b$ and $c$ quark masses minus a small correction for
binding).  One then predicts \cite{EHQ} nonstrange $B^{**}$ states with $J =
(1,2)$ at (5755, 5767) MeV, to which we shall return in Sec.~X A.
\bigskip

\leftline{\bf E.  Lifetime differences}
\bigskip

Charmed particle lifetimes range over a factor of ten, with
\begin{equation}
\tau(\Xi_c^0) < \tau(\Lambda_c) < \tau(\Xi_c) \simeq \tau(D^0) \simeq
\tau(D_s) < \tau(D^+)~~~.
\end{equation}

Effects which contribute to these differences \cite{lifes} include (a) an
overall nonleptonic enhancement from QCD \cite{enh}, (b) interference when at
least two quarks in the final state are the same \cite{int}, (c) exchange and
annihilation graphs, e.g. in $\Lambda_c$ and $\Xi_c^0$ decays \cite{exch}, and
(d) final-state interactions \cite{fsi}.

In the case of $B$ hadrons, theorists estimate that all these effects shrink in
importance to less than ten percent \cite{blife}.  However, since the measured
semileptonic branching ratio for $B$ decays of about 10 or 11\% differs from
theoretical calculations of 13\% by some 20\%, one could easily expect such
differences among different $b$-flavored hadrons.  These could arise, for
example, from final-state interaction effects.  There are many tests for such
effects possible in the study of decays of $B$ mesons to pairs of
pseudoscalars \cite{BPP}.
\bigskip

\leftline{\bf F.  Anomalous electroweak couplings of charm?}
\bigskip

A curious item was reported \cite{RB} at the DPF 94 conference in August in
Albuquerque.  The forward-backward asymmetries in heavy-quark production,
$A^{0,b}_{FB}$ and $A^{0,c}_{FB}$, have been measured both on the $Z$ peak and
2 GeV above and below it.  All quantities are in accord with standard model
expectations except for $A^{0,c}_{FB}$ at $M_Z - 2$ GeV, which is considerably
more negative than expected.  It would be interesting to see if this effect is
confirmed by other groups.
\bigskip

\centerline{\bf VII.  HEAVY QUARK SYMMETRY}
\bigskip

In a hadron containing a single heavy quark, that quark ($Q = c$ or $b$) plays
the role of an atomic nucleus, with the light degrees of freedom (quarks,
antiquarks, gluons) analogous to the electron cloud.  The properties of hadrons
containing $b$ quarks then can calculated from the corresponding properties of
charmed particles by taking account \cite{HQS} of a few simple ``isotope
effects.''  For example, if $q$ denotes a light antiquark, the mass of a $Q
\bar q$ meson can be expressed as
\begin{equation}
M(Q \bar q) = m_Q + {\rm const.}[n,L] + \frac{\langle p^2 \rangle}{2 m_Q} +
a \frac{\langle {\bf \sigma_q \cdot \sigma_Q} \rangle}{m_q m_Q} + {\cal O}
(m_Q^{-2})~~~.
\end{equation}
Here the constant depends only on the radial and orbital quantum numbers $n$
and $L$.  The $\langle p^2 \rangle /2m_Q$ term expresses the dependence of
the heavy quark's kinetic energy on $m_Q$, while the last term is a hyperfine
interaction.  The expectation value of $\langle {\bf \sigma_q \cdot \sigma_Q}
\rangle$ is $(+1,~-3)$ for $J^P = (1^-,~0^-)$ mesons. If we define
$\overline{M} \equiv [3 M(1^-) + M(0^-)]/4$, we find
\begin{equation}
m_b - m_c + \frac{\langle p^2 \rangle}{2 m_b} - \frac{\langle p^2 \rangle}
{2 m_c} = \overline{M}(B \bar q) - \overline{M}(c \bar q) \simeq 3.34~{\rm
GeV}~~.
\end{equation}
so $m_b - m_c > 3.34$ GeV, since $\langle p^2 \rangle > 0$.  Details of
interest include (1) the effects of replacing nonstrange quarks with strange
ones, (2) the energies associated with orbital excitations, (3) the size of the
$\langle p^2 \rangle$ term, and (4) the magnitude of hyperfine effects.  In all
cases there exist ways of using information about charmed hadrons to predict
the properties of the corresponding $B$ hadrons. In search of methods without
theoretical bias, we have even resorted \cite{KRb} on occasion to numerical
interpolation!
\newpage

\centerline{\bf VIII.  OVERVIEW OF HADRONS WITH BEAUTY}
\bigskip

The use of heavy quark symmetry allows us to extrapolate from the spectrum
shown in Fig.~7 of hadrons containing a single charmed quark to that of hadrons
containing a single $b$ quark.  Taking account of the effects mentioned in the
previous section, we obtain the spectrum shown in Fig.~10, updated and adapted
from Ref.~\cite{KQR}.  Some similarities and differences with respect to the
charmed-hadron spectrum can be seen.

\begin{figure}
\centerline{\epsfysize=4.7in \epsffile{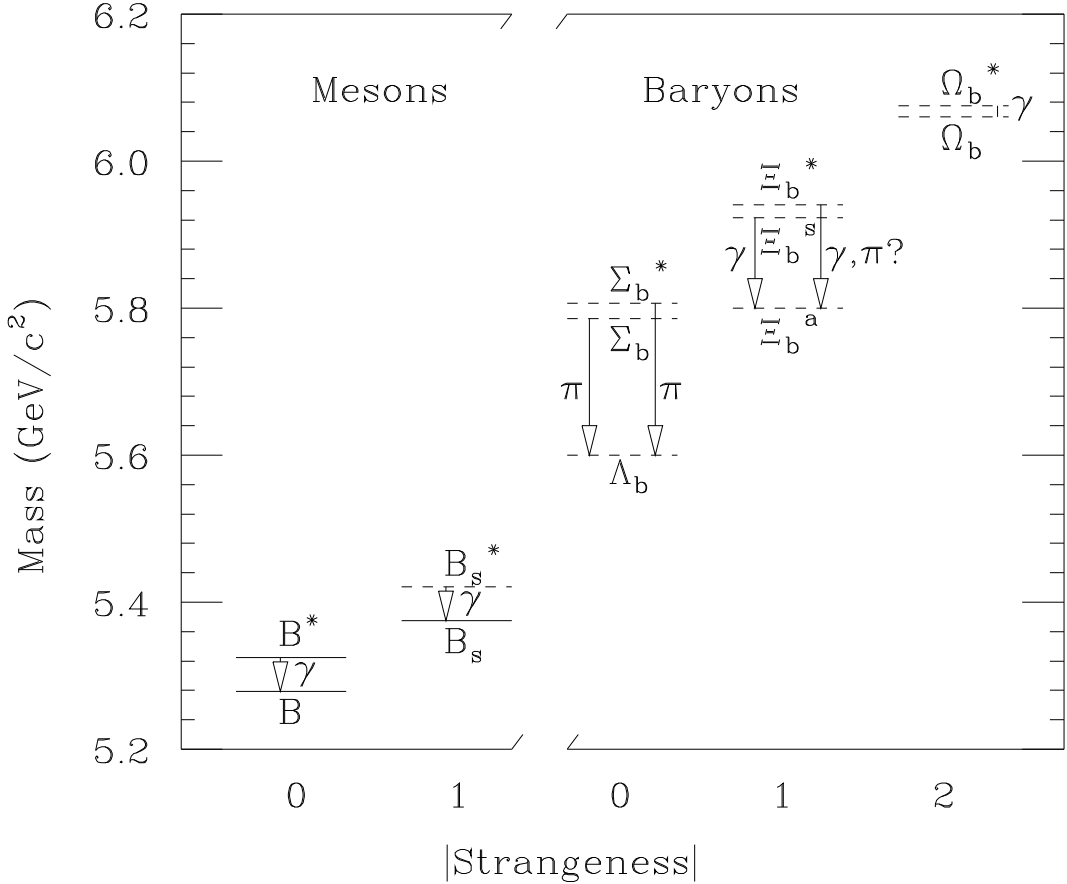}}
\caption{Lowest-lying S-wave levels of hadrons containing a single $b$
quark.  The lowest level in each group decays weakly.  Dashed lines indicate
levels not yet observed.}
\end{figure}

The $B^* - B$ hyperfine splitting scales as the inverse of the heavy-quark
mass: $B^* - B = (m_c/m_b)(D^* - D)$.  Consequently, while $D^{*+} \to D^0
\pi^+$ and $D^{*+} \to D^+ \pi^0$ are both allowed, leading to a useful
method \cite{SN} for identifying charmed mesons via the soft pions often
accompanying them, the only allowed decay of a $B^*$ is to $B \gamma$.  No soft
pions are expected to accompany $B$ mesons.

The $B_s^* - B_s$ hyperfine splitting is expected to be the same as that
between $B^{*0}$ and $B^0$ \cite{RW}, as mentioned earlier.  The observation by
the CUSB group \cite{CUSBBs} consistent with this expectation needs
confirmation.

In the $\Lambda_b$, the $u$ and $d$ quarks are in a state of zero spin and
isospin, so the $b$ quark carries the spin of the $\Lambda_b$.  This fact may
be useful in probing the weak interactions of the $b$ quark \cite{Aetal}.
Although the $\Lambda_b$ is listed as established by the Particle Data Group
\cite{PDG} (see the experiments in Ref.~\cite{LBEX}, yielding an
average mass of $5641 \pm 50$ MeV), its confirmation in Fermilab \cite{CDFLB}
and LEP experiments has remained elusive up to now. Bounds on its mass were
derived some time ago by Martin \cite{MartinL} and refined by Martin and
Richard \cite{MRL}.

Many other states are expected to be rather similar to those in the charm
system, once the added mass of the $b$ quark has been taken into account.
The precise value of the splitting between the $\Sigma_b^*$ and $\Sigma_b$
is important \cite{Sigb} in estimating the amount of depolarization
undergone by a $b$ quark as it fragments into a $\Lambda_b$.
\bigskip

\centerline{\bf IX.  THE CKM MATRIX}
\bigskip

Our present understanding of CP violation links the observed effect in the
neutral kaon system to a phase in the unitary Cabibbo-Kobayashi-Maskawa
\cite{CKM} (CKM) matrix describing weak charge-changing transitions among
quarks.  A sound understanding of the way in which heavy quarks are
incorporated into hadrons is essential to specify the CKM parameters precisely
as possible in order to test the theory.
\bigskip

\leftline{\bf A.  Measuring CKM elements}
\bigskip

We write the matrix in the form \cite{wp}:
\begin{equation}
V = \left ( \begin{array}{c c c}
V_{ud} & V_{us} & V_{ub} \\
V_{cd} & V_{cs} & V_{cb} \\
V_{td} & V_{ts} & V_{tb}
\end{array} \right )
\approx \left [ \matrix{1 - \lambda^2 /2 & \lambda & A \lambda^3 ( \rho -
i \eta ) \cr
- \lambda & 1 - \lambda^2 /2 & A \lambda^2 \cr
A \lambda^3 ( 1 - \rho - i \eta ) & - A \lambda^2 & 1 \cr } \right ]~~~~~ .
\end{equation}
The upper left $2 \times 2$ submatrix involves only one real parameter $\lambda
= \sin \theta_c$, where $\theta_c$ is the Cabibbo angle. The couplings
involving the third family of quarks $(b,t)$ require three additional
parameters $A,~\rho$, and $\eta$. We outline the means \cite{JRCKM} by which
these quantities are measured.

{\it 1.  The parameter $\lambda$} is specified by comparing strange
particle decays with muon decay and nuclear beta decay, leading to $\lambda
\approx \sin \theta \approx 0.22$.

{\it 2.  The element $V_{cb} = A \lambda^2$} is responsible for the dominant
decays of $b$-flavored hadrons.  The lifetimes of these hadrons and their
semileptonic branching ratios then lead to an estimate $V_{cb} = 0.038 \pm
0.003$, or $A = 0.79 \pm 0.06$.  One must relate processes at the quark level
to those at the hadron level either using a QCD-corrected free quark
estimate or specific models for final states.  The constraints on
$m_b - m_c$ arising in charmonium and upsilon spectroscopy, whereby this
difference lies between 3.34 and 3.4 GeV, are proving useful in this regard.

{\it 3.  The magnitude of the element $V_{ub}$} governs the rate of decays of
$b$-flavored hadrons to charmless final states.  One infers
$|V_{ub}/V_{cb}| = 0.08 \pm 0.02$ or $\sqrt{\rho^2 + \eta^2} = 0.36 \pm 0.09$
from leptons emitted in semileptonic decays $b \to u \ell
\nu$ with energies beyond the endpoint for $b \to c \ell \nu$.
The error reflects the uncertainty associated
with models relating this small part of the spectrum to the whole rate.

{ \it 4.  The phase of $V_{ub}$}, Arg $(V_{ub}^*) = \arctan(\eta/\rho)$, is the
least certain quantity.  Information on it may be obtained by studying its
effect on contributions of higher-order diagrams involving the top quark, such
as those governing $B^0 - \bar B^0$ mixing and CP-violating $K^0 - \bar K^0$
mixing,  with \cite{CDFtop} $m_t = 174 \pm 17$ GeV.

The most recent estimate for the $B^0 - \bar {B}^0$ mixing amplitude,
incorporating recent observations of time-dependent oscillations \cite{Sharma},
is $\Delta m/\Gamma = 0.71 \pm 0.07$.  The dominant contribution to the mixing
is provided by one-loop diagrams (``box graphs'') involving internal
$W$ and top quark lines, leading to $\Delta m \sim f_B^2 m_t^2 |V_{td}|^2$
(times a slowly varying function of $m_t/M_W$). Here the ``$B$ decay
constant,'' $f_B$, describes the amplitude for finding a $b$ antiquark and a
light quark at the same point in a $B$ meson.  Since $|V_{td}| \sim | 1 - \rho
- i \eta|$, the $B^0 - \bar {B}^0$ mixing amplitude leads to a constraint
in the $(\rho,\eta)$ plane consisting of a circular band with center (1,0).
The main contribution to the width of this band is uncertainty in $f_B$.

A similar set of box diagrams contributes to the parameter $\epsilon$
describing CP-violating $K^0 - \bar {K}^0$ mixing.  The imaginary part of
the mass matrix is proportional to $f_K^2 m_t^2 {\rm Im} (V_{td}^2)$ times a
slowly varying function of $m_t$, with a small correction for the charmed quark
contribution and an overall factor $B_K$ describing the degree to which the box
graphs account for the effect.  Since Im($V_{td}^2) \sim \eta (1 - \rho)$, the
constraint imposed by CP-violating $K^0 - \bar {K}^0$ mixing consists of a
hyperbolic band in the $(\rho,\eta)$ plane with focus at (1,0), whose width is
dominated by uncertainty in the magnitude of $V_{cb}$ \cite{StonePas}.
\bigskip

\leftline{\bf B.  Constraints on parameters}
\bigskip

The allowed region in the $(\rho,\eta)$ plane is bounded by circular bands
associated with the $|V_{ub}/V_{cb}|$ and $B^0 - \bar B^0$ mixing constraints,
and a hyperbolic band associated with the CP-violating $K^0 - \bar K^0$ mixing
constraint.  In a recent determination \cite{DPF} we used parameters, in
addition to those mentioned above, including $B_K = 0.8 \pm 0.2$, $f_B = 180
\pm 30$ MeV (in units where $f_\pi = 132$ MeV), $\eta_{\rm QCD} = 0.6 \pm 0.1$
(a correction to the $B - \bar {B}$ mixing diagrams), and $B_B = 1$ for the
factor analogous to $B_K$, and found $-0.3 \le \rho \le 0.3$, $0.2 \le \eta \le
0.4$.  The main uncertainty in $\rho$ stems from that in $f_B$, while
model-dependent sources of error in $V_{cb}$ and $V_{ub}$ are the main sources
of uncertainty on $\eta$.  Thus, improved knowledge about hadron physics can
have a major impact on our present understanding of weak interactions.
\bigskip

\leftline{\bf C.  $B_s - \bar {B}_s$ mixing}
\bigskip

In contrast to $B^0 - \bar {B}^0$ mixing, which involves the uncertain
CKM element $V_{td}$, the $B_s - \bar {B}_s$ mixing amplitude involves
the elements $V_{ts} \approx - V_{cb} = -0.038 \pm 0.003$ and $V_{tb} \approx
1$, so that the main source of uncertainty in $x_s \equiv (\Delta
m/\Gamma)_{B_s}$ is the decay constant $f_{B_s}$.  For $f_{B_s} = 200 \pm 50$
MeV and $m_t = 174 \pm 17$ GeV we find \cite{DPF} $x_s = 16 \times
2^{\pm 1}$. If this mixing rate can be measured and the
uncertainties on $V_{ts}$ and $m_t$ reduced, a useful value for
$f_{B_s}$ (and hence, via SU(3), for $f_B$) can be obtained.  Estimates for
$f_B/f_{B_s}$ range from about 0.8 to 0.9 \cite{DPF}.
\bigskip

\centerline{\bf X.  CP VIOLATION IN $B$ DECAYS}
\bigskip

If the phase in the CKM matrix is responsible for CP violation in the
neutral kaon system, dramatic CP-violating effects are expected in decays
of $B$ mesons.  In order to exploit and interpret these effects, many
aspects of hadron spectroscopy are important.  I would like to mention just
two areas of recent progress.
\bigskip

\leftline{\bf A.  Decays to CP eigenstates}
\bigskip

{\it 1.  $\pi - B$ correlations} are useful in identifying the flavor of
neutral $B$ mesons at the time of production.  Once produced, these mesons
can undergo $B^0 - \bar {B}^0$ mixing, leading to time-dependent
asymmetries in decays to CP eigenstates like $J/\psi K_S$.  Time-integrated
decays also can display rate asymmetries, whose interpretation is often
independent of final-state effects.  For example, the asymmetry in decays of
$B^0$ or $\bar {B}^0$ to $J/\psi K_S$ is equal to $-[x_d/(1 + x_d^2)]\sin
[{\rm Arg} (V_{td}^*)^2]$, where $x_d = (\Delta m / \Gamma)|_d = 0.70 \pm 0.07$
is the mixing parameter mentioned earlier.  One has to know the flavor of the
neutral $B$ at time of production.  One proposed means for ``tagging'' the
$B$ involves its correlation with charged pions produced nearby in phase
space \cite{GNR}.  The existence of such a correlation is predicted both by
fragmentation and resonance decay pictures.

{\it 2.  $B^{**}$ resonances} can serve as explicit sources of $\pi - B$
correlations.  One expects resonances in the $\pi^+ B^0$ and $\pi^-
\bar {B}^0$ channels but not in the $\pi^- B^0$ and $\pi^+ \bar {B}^0$
channels.  If these resonances are narrow, they can help in suppressing
backgrounds.

The expected spectrum of $B^{**}$ resonances can be roughly anticipated by
adding about 3.32 GeV to the masses of excited charmed states shown in Fig.~9.
One expects narrow P-wave levels of spins 1 and 2 around 5.76 GeV, and
broader levels of spins 0 and 1 somewhat lower in mass.  Recently two groups
at LEP \cite{OPALDELPHI} have presented evidence for $\pi - B$ correlations
which appear to show at least some resonant activity in the ``right-sign''
combinations.
\bigskip

\leftline{\bf B.  Decays to CP non-eigenstates}
\bigskip

A difference between the rates for a process and its charge-conjugate, such as
$B^+ \to \pi^+ K^0$ and $B^- \to \pi^- K^0$, signifies CP violation.  Under
charge conjugation, weak phases change sign, but strong phases do not.  In
order for a rate difference to appear, there must be both a weak phase
difference and a strong phase difference in the channels with isospins $I =
1/2$ and 3/2.  Recently it has been shown that one may be able to measure weak
phases via the rates for $B$ decays to pairs of light pseudoscalar mesons {\it
without} having any strong phase differences \cite{BPP}.  The presence of
electroweak penguins \cite{DH} is one possible obstacle to this program, which
is under further investigation.
\bigskip

\centerline{\bf XI.  FOR THE FUTURE}
\bigskip

\leftline{\bf A.  Charmonium}
\bigskip

The study of charmonium levels not limited to those with $J^{PC} = 1^{--}$ will
benefit from further experiments with stored antiprotons \cite{E760}. One can
look forward to discovery of the $\eta_c'$, the narrow $1^{1,3}D_2$ levels, and
perhaps one or more narrow $2P$ levels.  The Beijing Electron-Positron Collider
will turn its attention to the $\psi(3770)$, a copious source of $D \bar D$
pairs, leading to an eventual measurement of the valuable $D$ meson decay
constant when the process $D \to \mu \nu$ is seen.
\bigskip

\leftline{\bf B.  Upsilons}
\bigskip

A number of interesting states remain to be found.  These include the $\eta_b$
(probably reachable from the $\Upsilon(2S)$), the $\eta_b'$, the lowest $^1P_1$
level (around 9.9 GeV), and various $\Upsilon(1D)$ and $\Upsilon(2D)$ states.
A careful scan in $e^+ e^-$ center-of-mass energy around 10.16 and 10.44 GeV
may be able to turn up the predicted $^3D_1$ levels.
\bigskip

\leftline{\bf C.  Charmed hadrons}
\bigskip

We can look forward to more precise measurements of the $D_s$ decay constant
and to the first observations of $D \to \mu \nu$.  The $\tau \nu$ final state
may be the largest single decay mode of the $D_s$, with a branching ratio
approaching 9\%!

The $2S$ charmed hadrons are expected to have masses of around 2.7 GeV,
and thus to be able to decay to $D_s^{(*)} \bar K$.  The discovery of such
modes would encourage us to look for similar correlations in $B_s K$ systems,
which would be useful in identifying the flavor of strange $B$ mesons at time
of production \cite{AB}.

Great progress has already been made, and more is expected, in the study of
charmed baryons (both S-wave and P-wave) and of P-wave charmed mesons. We can
look forward to the eventual discovery of charmed baryons with spins of 3/2,
the partners of the familiar $\Delta$ and $\Omega^-$.  The interest in the
masses and decays of these states transcends the charm sector alone, and is
important for anticipating properties of baryons containing a single $b$ quark.

The differences in charmed particle lifetimes have provided a wealth of
information about how strong interactions affect weak decays.  These
differences are expected to be much less marked for hadrons with beauty.  One
baryon whose lifetime is expected to be very short \cite{lifes} is the
$\Omega_c$; we look forward to a determination (or at least an upper limit) in
the near future.

Hadrons with more than one charmed quark (such as the $ccu$ baryon) provide an
interesting testing ground for theorems concerning the masses of multi-quark
systems \cite{multi}.  Perhaps such hadrons can be produced in $e^+ e^-$
interactions, where one does not have to pay a heavy penalty for production
of the first charmed quark.
\bigskip

\leftline{\bf D.  Hadrons with beauty}
\bigskip

In a few years, we will have confirmed the existence of the $B_s^*$, the
$\Lambda_b$, the narrow $1P$ mesons, and perhaps some $2S$ states as well.
The $1P$ mesons in particular may prove a valuable adjunct to CP-violation
studies in the $B$ meson system.

A great deal remains to be learned about the weak decays of hadrons
with beauty, especially to charmless final states.  Here experimental work has
outstripped theory in most cases, requiring us to come up with more reliable
models for the way in which quarks are incorporated into hadrons.  One area
of future experimental progress may be in the determination of the full
kinematics of semileptonic decay processes (including the momentum of the
missing neutrino), which will reduce dependence on models.

With luck and ingenuity, we may yet learn the amplitude for $B_s -
\bar {B}_s$ mixing, which will help fix the decay constant $f_{B_s}$
and, via SU(3), the constant $f_B$ which is of great importance in
anticipating CP-violating effects in the $B$ meson system.

Finally, we can look forward to many years of fine data from CESR, Fermilab,
LEP, and future facilities, culminating in the discovery of $CP$ violation in
$B$ decays.  This would represent a triumph of Standard Model physics based on
our present picture of the CKM matrix. We would then have to figure out where
that curious phase in the CKM matrix actually comes from!
\bigskip

\leftline{\bf E.  Conclusion}
\bigskip

In conclusion, let me express thanks on behalf of all of us at this symposium
to Andr\'e Martin for showing us physics with charm and beauty!
\bigskip

\centerline{\bf ACKNOWLEDGMENTS}
\bigskip

This article is dedicated to the memory of M. A. Baqi B\'eg, whose kindness
and gentle advice I have appreciated since my earliest days in particle
physics. I am grateful to Andr\'e Martin for the opportunity to present this
review and to fruitful correspondence and discussions over the years.  In
addition to him, many people have contributed to the work reported here,
including J. L. Basdevant, J. Amundson, B. Baumgartner, R. Bertlmann, A.
Common, I. Dunietz, A. Grant, H. Grosse, W. Kwong, H. Lipkin, P. Moxhay, C.
Quigg, J.-M. Richard, H. Riggs, E. Rynes, J. Schonfeld, J. Stubbe, P. Taxil, H.
Thacker, M. Wise, and T. T. Wu. I wish to thank the CERN Theory Group for its
hospitality.  This work was supported in part by the United States Department
of Energy under Contract No. DE FG02 90ER40560.

\def \ajp#1#2#3{Am.~J.~Phys.~{\bf#1}, #2 (#3)}
\def \apny#1#2#3{Ann.~Phys.~(N.Y.) {\bf#1}, #2 (#3)}
\def \app#1#2#3{Acta Phys.~Polonica {\bf#1}, #2 (#3)}
\def \arnps#1#2#3{Ann.~Rev.~Nucl.~Part.~Sci.~{\bf#1}, #2 (#3)}
\def \cmp#1#2#3{Commun.~Math.~Phys.~{\bf#1}, #2 (#3)}
\def \cmts#1#2#3{Comments on Nucl.~Part.~Phys.~{\bf#1}, #2 (#3)}
\def \cn{Collaboration}
\def \corn93{{\it Lepton and Photon Interactions:  XVI International Symposium,
Ithaca, NY August 1993}, AIP Conference Proceedings No.~302, ed.~by P. Drell
and D. Rubin (AIP, New York, 1994)}
\def \cp89{{\it CP Violation,} edited by C. Jarlskog (World Scientific,
Singapore, 1989)}
\def \dpf94{DPF 94 Meeting, Albuquerque, NM, Aug.~2--6, 1994}.
\def \efi{Enrico Fermi Institute Report No. EFI}
\def \el#1#2#3{Europhys.~Lett.~{\bf#1}, #2 (#3)}
\def \f79{{\it Proceedings of the 1979 International Symposium on Lepton and
Photon Interactions at High Energies,} Fermilab, August 23-29, 1979, ed.~by
T. B. W. Kirk and H. D. I. Abarbanel (Fermi National Accelerator Laboratory,
Batavia, IL, 1979}
\def \hb87{{\it Proceeding of the 1987 International Symposium on Lepton and
Photon Interactions at High Energies,} Hamburg, 1987, ed.~by W. Bartel
and R. R\"uckl (Nucl. Phys. B, Proc. Suppl., vol. 3) (North-Holland,
Amsterdam, 1988)}
\def \ib{{\it ibid.}~}
\def \ibj#1#2#3{~{\bf#1}, #2 (#3)}
\def \ichep72{{\it Proceedings of the XVI International Conference on High
Energy Physics}, Chicago and Batavia, Illinois, Sept. 6--13, 1972,
edited by J. D. Jackson, A. Roberts, and R. Donaldson (Fermilab, Batavia,
IL, 1972)}
\def \ijmpa#1#2#3{Int.~J.~Mod.~Phys.~A {\bf#1}, #2 (#3)}
\def \ite{{\it et al.}}
\def \jmp#1#2#3{J.~Math.~Phys.~{\bf#1}, #2 (#3)}
\def \jpg#1#2#3{J.~Phys.~G {\bf#1}, #2 (#3)}
\def \lkl87{{\it Selected Topics in Electroweak Interactions} (Proceedings of
the Second Lake Louise Institute on New Frontiers in Particle Physics, 15--21
February, 1987), edited by J. M. Cameron \ite~(World Scientific, Singapore,
1987)}
\def \ky85{{\it Proceedings of the International Symposium on Lepton and
Photon Interactions at High Energy,} Kyoto, Aug.~19-24, 1985, edited by M.
Konuma and K. Takahashi (Kyoto Univ., Kyoto, 1985)}
\def \mpla#1#2#3{Mod.~Phys.~Lett.~A {\bf#1}, #2 (#3)}
\def \nc#1#2#3{Nuovo Cim.~{\bf#1}, #2 (#3)}
\def \np#1#2#3{Nucl.~Phys.~{\bf#1}, #2 (#3)}
\def \pisma#1#2#3#4{Pis'ma Zh.~Eksp.~Teor.~Fiz.~{\bf#1}, #2 (#3) [JETP Lett.
{\bf#1}, #4 (#3)]}
\def \pl#1#2#3{Phys.~Lett.~{\bf#1}, #2 (#3)}
\def \plb#1#2#3{Phys.~Lett.~B {\bf#1}, #2 (#3)}
\def \pr#1#2#3{Phys.~Rev.~{\bf#1}, #2 (#3)}
\def \pra#1#2#3{Phys.~Rev.~A {\bf#1}, #2 (#3)}
\def \prd#1#2#3{Phys.~Rev.~D {\bf#1}, #2 (#3)}
\def \prl#1#2#3{Phys.~Rev.~Lett.~{\bf#1}, #2 (#3)}
\def \prp#1#2#3{Phys.~Rep.~{\bf#1}, #2 (#3)}
\def \ptp#1#2#3{Prog.~Theor.~Phys.~{\bf#1}, #2 (#3)}
\def \rmp#1#2#3{Rev.~Mod.~Phys.~{\bf#1}, #2 (#3)}
\def \rp#1{~~~~~\ldots\ldots{\rm rp~}{#1}~~~~~}
\def \si90{25th International Conference on High Energy Physics, Singapore,
Aug. 2-8, 1990}
\def \slc87{{\it Proceedings of the Salt Lake City Meeting} (Division of
Particles and Fields, American Physical Society, Salt Lake City, Utah, 1987),
ed.~by C. DeTar and J. S. Ball (World Scientific, Singapore, 1987)}
\def \slac89{{\it Proceedings of the XIVth International Symposium on
Lepton and Photon Interactions,} Stanford, California, 1989, edited by M.
Riordan (World Scientific, Singapore, 1990)}
\def \smass82{{\it Proceedings of the 1982 DPF Summer Study on Elementary
Particle Physics and Future Facilities}, Snowmass, Colorado, edited by R.
Donaldson, R. Gustafson, and F. Paige (World Scientific, Singapore, 1982)}
\def \smass90{{\it Research Directions for the Decade} (Proceedings of the
1990 Summer Study on High Energy Physics, June 25 -- July 13, Snowmass,
Colorado), edited by E. L. Berger (World Scientific, Singapore, 1992)}
\def \stone{{\it B Decays}, edited by S. Stone (World Scientific, Singapore,
1994)}
\def \tasi90{{\it Testing the Standard Model} (Proceedings of the 1990
Theoretical Advanced Study Institute in Elementary Particle Physics, Boulder,
Colorado, 3--27 June, 1990), edited by M. Cveti\v{c} and P. Langacker
(World Scientific, Singapore, 1991)}
\def \yaf#1#2#3#4{Yad.~Fiz.~{\bf#1}, #2 (#3) [Sov.~J.~Nucl.~Phys.~{\bf #1},
#4 (#3)]}
\def \zhetf#1#2#3#4#5#6{Zh.~Eksp.~Teor.~Fiz.~{\bf #1}, #2 (#3) [Sov.~Phys. -
JETP {\bf #4}, #5 (#6)]}
\def \zpc#1#2#3{Zeit.~Phys.~C {\bf#1}, #2 (#3)}

\end{document}